\documentclass[a4paper,twocolumn,prl,nofootinbib,floatfix,showpacs]{revtex4}

\usepackage{amsmath}
\usepackage{amssymb} 
\usepackage{epstopdf}
\usepackage{graphicx}

\newcommand{\bra}[1]{\langle #1 |} 
\newcommand{\ket}[1]{| #1 \rangle } 
\newcommand{\braket}[2] {\langle #1 | #2\rangle } 
\newcommand{\braoket}[3] {\langle #1 | #2 | #3 \rangle } 
\newcommand{\norm}[1] {|| #1 ||} 
\newcommand{\expect}{\mathbb{E}} 

\begin{document}

\title{Convergence of repeated quantum non-demolition measurements\\ and wave function
  collapse}

\author{Michel Bauer ${}^{\spadesuit,\clubsuit~}$\cite{michel} and Denis Bernard
  ${}^{\clubsuit~}$\cite{denis}}

\date{\today} 

\affiliation{ ${}^\spadesuit$ Institut de Physique Th\'eorique de Saclay
  \cite{ipht}, 
  CEA-Saclay, 91191 Gif-sur-Yvette, France.\\
  $^\clubsuit$
  Laboratoire de Physique Th\'eorique de l'Ecole Normale Sup\'erieure, \\
  CNRS/ENS, Ecole Normale Sup\'erieure, 24 rue Lhomond, 75005 Paris, France }

\preprint{ Preprint IPhT 2011/??? ; arxiv?/??? }
        
\begin{abstract}
Motivated by recent experiments on quantum trapped fields, we give a rigorous proof that repeated indirect quantum non-demolition (QND) measurements converge to the collapse of the wave function as predicted by the postulates of quantum mechanics for direct measurements. We also relate the
rate of convergence towards the collapsed wave function to the relative entropy
of each indirect measurement, a result which makes contact with information theory.  
\end{abstract}


\maketitle

Wave  function collapse is a basic axiom of quantum direct measurement \`a la Von Neumann \cite{mes}. A quantum non-demolition measurement \cite{XXX} is one for which the collapsed state is an eigenstate of the free evolution. Repeating the measurement on the collapsed state yields identical results since this state is preserved by the evolution. Indirect measurements \cite{YYY} consists in letting the quantum system under study be entangled with another quantum system, called the probe, and in implementing a direct measurement on the probe. Since the system and the probe are entangled, one gains information. Repeating the process of entanglement and measurement increases statistically the information one gets on the system.

Developing experimental and theoretical expertise on quantum measurement processes is mandatory for developing quantum state manipulation.
It was early realized \cite{Davies,Hepp} that modeling quantum measurements require systems with infinitely many degrees of freedom, e.g. as in the phenomenological stochastic models of \cite{Gisin-Diosi}. The need to describe quantum jumps and randomness inherent to repeated measurements lead to the concept of quantum trajectories \cite{DCM,Charmichael}. In parallel, tools of open quantum systems, specifically those of quantum stochastic calculus \cite{HP}, have been adapted to the description of quantum continual measurements \cite{GBB} and quantum feedback \cite{Wiseman}. In most of these stochastic models, the driving noises, often classical or quantum Brownian motions, are linked to the degrees of freedom of the measurement apparatus. 
Although bearing similarities with these frameworks, our proof of the wave collapse in series of QND measurements is based on a purely quantum description of the repeated probe-system interactions.

\begin{figure}
\mbox{\includegraphics[width=0.48\textwidth]{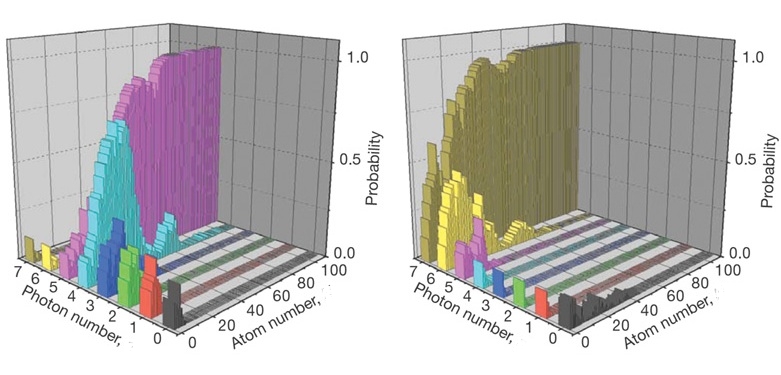}}
\caption{\emph{Two experimental samples \cite{credit_fig} of reconstructed photon distributions as functions of the number of indirected measurements (i.e. the number of atoms traversing the cavity) according to \cite{lkb_exp}. The collapse of the photon distribution to a realization dependent sharply defined number is clearly visible.}}
\label{fig:lkb_exp}
\end{figure}

Experiments on repeated indirect quantum non-demolition measurements have recently been performed, in particular in quantum optics. As an example, let us look at \cite{lkb_exp} whose setup is the following. The tested quantum system is a resonant electromagnetic cavity selecting photons of given frequency. It is probed by sending Rydberg atoms through it, one after the other. During the atom-photon interaction each atom behaves as a two-state system modeled by a spin one-half \cite{rydberg}. The atoms are prepared with their effective spins pointing in the $0x$ direction \cite{0xyz}. The experimental protocol ensures that the atom effective spin rotates around the $0z$ axis by an angle proportional to the number of photons $\hat n_{\rm ph}$ in the cavity, say $\hat n_{\rm ph}\theta$ with $\theta$ a fixed angle. After interaction, the atom-photon system is entangled, but the cavity state gets unchanged if it is initially an eigenstate of the free photon hamiltonian. The effective atom spin is then measured along a direction perpendicular to $0z$ but at angle $\phi$ with respect to $0x$. The output of the spin measure is $\pm$ with probabilities $p_+(\phi|\hat n_{\rm ph})=\cos^2[{(\hat n_{\rm ph}\theta-\phi)}/{2}]$ and $p_-(\phi|\hat n_{\rm ph})=\sin^2[{(\hat n_{\rm ph}\theta-\phi)}/{2}]$, if there are $\hat n_{\rm ph}$ photons in the cavity. If the initial photon distribution is $q_0(\hat n_{\rm ph})$, the probability to measure an effective spin $\pm$ is $\sum_{\hat n_{\rm ph}}q_0(\hat n_{\rm ph})p_\pm(\phi|\hat n_{\rm ph})$. No direct measurement on the cavity is done. The experimental aim is to reconstruct the initial photon distribution by accumulating informations from the repeated atom effective spin measurements. The photon distribution is recalculated after each atom measurement using Bayes law \cite{bayes}. Fig.\ref{fig:lkb_exp} shows experimental data for the evolution of reconstructed photon distributions. For each realization, they converge, experimentally and numerically \cite{lkb_exp}, to peaked distributions whose centers depend on the realization. This is the collapse.

Let us abstract and generalize the previous situation.
At initial time, the system is in state $\ket{\varphi _0}\equiv\ket{\varphi}$. It
interacts during time $\Delta t$ with a probe initially in state
$\ket{\psi}$, so that the pair (probe$+$system) evolves into $U (\ket{\psi}
\otimes \ket{\varphi})$, where $U$ is some unitary operator acting on the Hilbert space ${\mathcal
  H}_{probe} \otimes {\mathcal H}_{syst}$.  After $\Delta t$, the system-probe interaction can be neglected. 
A perfect measurement \`a la Von Neumann is then
performed on the probe. This means that there is an orthonormal basis $\ket{i}$,
$i\in I$, of ${\mathcal H}_{probe}$ such that, after the measurement, the 
(probe$+$system)-state is proportional to 
$(\ket{i}\bra{i}\otimes {\rm Id}) U (\ket{\psi} \otimes\ket{\varphi})$
with probability $\norm{(\ket{i}\bra{i}\otimes {\rm Id}) U (\ket{\psi} \otimes
  \ket{\varphi})}^2$. The vanishing of this probability for a certain state
$\ket{i}$ means that the probe cannot be found in state $\ket{i}$, so  we can 
(and shall) simply forget about that possibility. 

We make the following assumption, related to non-demolition, on the evolution operator $U$: there is an
orthonormal basis $\ket{\alpha}$, $\alpha \in A$, of ${\mathcal H}_{syst}$ and
a collection of operators $U_{\alpha}$ acting on ${\mathcal H}_{probe}$ such
that, for each $\alpha$, 
\begin{equation}  \label{eq:decomp}
U(\ket{\psi} \otimes
  \ket{\alpha})=(U_{\alpha}\ket{\psi})\otimes \ket{\alpha}.
\end{equation}
The operators $U_{\alpha}$ are automatically unitary. 
If the probe is found in state $\ket{i}$ after the measurement,
(probability $\sum_{\alpha \in A}|\braoket{i}{U_{\alpha}}{\psi}|^2
|\braket{\alpha}{\varphi}|^2$), the pair (probe$+$system) is again in a tensor
product state $\ket{i}\otimes \ket{\varphi_1}$ where
\begin{equation} \label{eq:vec}
  \ket{\varphi_1}=\frac{\sum_{\alpha \in A} \braoket{i}{U_{\alpha}}{\psi}
    \braket{\alpha}{\varphi}\ \ket{\alpha}}{\left(\sum_{\alpha \in
        A}|\braoket{i}{U_{\alpha}}{\psi}|^2
      |\braket{\alpha}{\varphi}|^2\right)^{1/2}}.
\end{equation}
It is clear that the motivating experiment fulfills this property if $\ket{\alpha}$
is the occupation number basis \cite{Ulkb_exp}. 

The physics of this hypothesis is that the final
aim is to measure an observable on the system for whom the states
$\ket{\alpha}$ are eigenstates. As we shall see,
this (direct) measurement can be (indirectly) achieved by repeated
measurements on successive probes. So, one presents another probe to the system in
state $\ket{\varphi_1}$, let them interact and, after interaction,
measures the probe to get $\ket{\varphi_2}$ and so on. Notice that, in general,
at each step one could change the probe initial state, the observable
measured on the probe (this is indeed what happens in the motivating example),
and even the type of probes: the only thing one has to keep fixed is the
basis $\ket{\alpha}$ for which property (\ref{eq:decomp}) holds. Most of the
following discussion can be extended to the general setting \cite{mb_prep} but to keep notation
simple, we concentrate on the case when $\ket{\psi}$ and the
basis $\ket{i}$ are the same for all probes.\\

We start with a summary of our results: \\
$i)$ If a series of repeated indirect measurements is conducted, the state of the system
will stabilize over time and go to a limit. Carrying identical independent experiments
again, the system state will stabilize over time again but possibly with
different limits.\\ 
$ii)$ Under a physically meaningful non-degeneracy condition, the only possible
limits for the state of the system are the pointer states $\ket{\alpha}$, and
the probability to end in state $\ket{\alpha}$ starting from state
$\ket{\varphi}$ is $|\braket{\alpha}{\varphi}|^2$. Hence the outcome of a large number of
repeated indirect measurements satisfying condition (\ref{eq:decomp}) obeys the
standard rules of quantum mechanics direct measurements.\\
$iii)$ Under the same non-degeneracy condition, the measurements on the  probes 
allow to infer the limit pointer state for each independent experiment.\\
$iv)$ The rate of convergence to one of the pointer states is governed by the
relative entropy of certain probability measures in classical probe space.  The
order of magnitude of the probability that, while the repeated measurements are
conducted, the state of the system comes close to a pointer state but ends up
finally in another one can be computed explicitly.\\

The tools to prove these statements come from the classical theory of random processes : strong law of large numbers,
martingale convergence theorem, large deviations. A proof of the wave function collapse using the martingale convergence theorem appeared in  \cite{Adler}. These works are based on non-linear stochastic extensions of the Schr\"odinger equation \cite{limit} whereas our results are pure consequences of quantum mechanics (with measurements on probes)~\cite{point} and are closer in spirit to quantum trajectory approaches \cite{DCM,Charmichael} and to experiments.

We now turn to the proofs. One can rephrase eq.(\ref{eq:vec}) by saying that, for each $\alpha \in A$,  
\[\braket{\alpha}{\varphi_1}=\frac{\braoket{i}{U_{\alpha}}{\psi}
  \braket{\alpha}{\varphi_0}}{\left(\sum_{\alpha \in
      A}|\braoket{i}{U_{\alpha}}{\psi}|^2
    |\braket{\alpha}{\varphi_0}|^2\right)^{1/2}} \]
if the probe is found in state $\ket{i}$.
Thus, a crucial consequence of (\ref{eq:decomp}) is that there are no
interference terms for different $\alpha$'s, so that taking the modulus squared
does not lead to (much) loss of information. We set $p(i|\alpha)\equiv
|\braoket{i}{U_{\alpha}}{\psi}|^2$,  and $q_0(\alpha)\equiv
|\braket{\alpha}{\varphi_0}|^2$, $q_1(\alpha)\equiv
|\braket{\alpha}{\varphi_1}|^2$, $q_2(\alpha)\equiv
|\braket{\alpha}{\varphi_2}|^2$  and so on. Observe that after measuring
the $n^{\rm th}$ probe one has, for each $\alpha \in A$,
\begin{equation} \label{eq:rs}
q_{n+1}(\alpha)=q_{n}(\alpha)\frac{p(i|\alpha)}{\sum_{\beta \in
      A}q_{n}(\beta) p(i|\beta)}  
\end{equation}
with probability $\pi_n(i)\equiv \sum_{\beta \in A} q_{n}(\beta) p(i|\beta)$.
 
This is a random recursion relation which is of Markovian type : to compute
the possible values of $q_{n+1}(\alpha)$ and their respective probabilities, all
one needs to know are the $q_{n}(\beta)$'s. Each probe measurement 
leads to a choice among the probe states $|i\rangle$ such that $\pi_n(i)\neq
0$. The question to be settled is the long time
behavior of the resulting random sequences $q_{n}(\alpha)$. 

Observe that the $q_{n}(\alpha)$'s and the $p(i|\alpha)$'s are $\geq 0$.
Moreover, $\sum_{\alpha \in A}q_{n}(\alpha)=1$ and $\sum_{i\in I} p(i|\alpha)=1$
for each $\alpha \in A$. It follows that $\sum_{i\in I} \pi_n(i)=1$ as it should
be. A crucial question is the following : having observed the random
sequences $q_{m}(\beta)$ for $m=0,\cdots,n$ and all $\beta$'s in $A$, what is
the average value of $q_{n+1}(\alpha)$? From (\ref{eq:rs}), it is immediate that
this (conditional) average, which we denote by $\expect
(q_{n+1}(\alpha)|q_0,\cdots,q_n)$, is
\begin{eqnarray*}
  \expect(q_{n+1}(\alpha)|q_0,\cdots,q_n) & = & \sum_{i, \pi_n(i)\neq 0}
  q_{n}(\alpha)\frac{p(i|\alpha)}{\pi_n(i)}\pi_n(i) \\ & = & \sum_{i,
    \pi_n(i)\neq 0} q_{n}(\alpha) p(i|\alpha). 
\end{eqnarray*}
Now,  $\pi_n(i)=\sum_{\beta \in A} q_{n}(\beta) p(i|\beta)$ and for this to 
vanish, the product $q_{n}(\beta) p(i|\beta)$ has to vanish for all $\beta \in
A$, and in particular for $\beta=\alpha$, so that $\sum_{i,
    \pi_n(i)\neq 0} q_{n}(\alpha) p(i|\alpha)=\sum_{i \in I} q_{n}(\alpha)
  p(i|\alpha)=q_{n}(\alpha)$. Hence we find that
\begin{equation}
\expect(q_{n+1}(\alpha)|q_0,\cdots,q_n) = q_{n}(\alpha).
\end{equation}

In the theory of random processes, such a property defines the concept of
martingale : the sequence $q_{0},q_1,\cdots $ is a martingale, because if one
knows it up to time $n$ (i.e. if one knows $q_{0},\cdots ,q_n$) its expectation
at time $n+1$ is its value at time $n$ (i.e. $q_n$). 
To connect quantum measures to conditional expectations is 
not so surprising because both rely on orthogonal projections in Hilbert spaces.

The martingale at hand has a peculiar property: 
it is bounded (every $q_{n}(\alpha)$ is $\geq 0$
and $\sum_{\alpha \in A} q_{n}(\alpha)=1$). We can then quote a special case of
the martingale convergence theorem (see any modern textbook on probability theory, 
e.g \cite{martingconv}, for a precise statement):
\textit{A random sequence $q_{0},q_1,\cdots $ which is a bounded martingale
  converges almost surely and in ${\mathbb L}^1$.  The limit, a random variable
  $q_{\infty}$, is such that its expectation satisfies
  $\expect(q_{\infty})=q_{0}$.}

This is a deep theorem and there is no intuitive argument that we know
to explain it \cite{martin}. But in our case its meaning is simple. The
statement of almost sure convergence is precisely the mathematical formulation of
$i)$. The statement of ${\mathbb L}^1$ convergence is a simple consequence
of the Lebesgue dominated convergence theorem, because our martingale is
bounded. The statement on the expectation of the limit random variable
yields the second part of $ii)$ once we have given an independent argument to
show that the possible limits are the pointer states.

To get this, we observe that the convergence of $q_{n}(\alpha)$ leads to the
convergence of $\pi_n(i)=\sum_{\beta \in A} q_{n}(\beta) p(i|\beta)$. If $i$ is
such that $\pi_{\infty}(i)\neq 0$ then, for $n$ large enough, $\pi_n(i) >
\pi_{\infty}(i)/2 > 0$ which implies that, with probability $1$, the $n^{th}$ probe will
be found in state $i$ for arbitrarily large values of $n$. This allows to take the
large $n$ limit in (\ref{eq:rs}) for this value of $i$. Hence
\[
q_{\infty}(\alpha)=q_{\infty}(\alpha)\frac{p(i|\alpha)}{\sum_{\beta \in
      A}q_{\infty}(\beta) p(i|\beta)},
\]
for any $i\in I$ such that $\pi_{\infty}(i)\neq 0$. Only the $\alpha$'s for
which $q_{\infty}(\alpha)\neq 0$ yield a nontrivial equation, so we can restrict
to these $\alpha$'s. Then, we can simplify to get $p(i|\alpha)=\pi_{\infty}(i)$
for any $i$ such that $\pi_{\infty}(i)\neq 0$. If $q_{\infty}(\alpha)\neq 0$, $\pi_{\infty}(i)=0$ implies
$p(i|\alpha)=0$, so that $p(i|\alpha)=\pi_{\infty}(i)$ is actually valid
for any $i$.  The right-hand side may depend on $i$ but it does not depend on
$\alpha$. So the same holds for the left-hand side: this means that the
evolution operator $U$ and the probe measurement act in a degenerate way on the
corresponding kets $\ket{\alpha}$. In such a degenerate situation, we cannot
expect to measure them individually, just as in a standard quantum measure of a
system observable we cannot separate the $\ket{\alpha}$'s having the same
eigenvalue \cite{period}. So, we
assume that for any $\alpha,\beta \in A$ there is some $i\in I$ such that
$p(i|\alpha)\neq p(i|\beta)$, and we get that
$q_{\infty}(\alpha)=\delta_{\alpha,\gamma}$ for some $\gamma$, i.e. the only
possible values for $q_{\infty}(\alpha)$ are $0$ or $1$. The equality
$\expect(q_{\infty}(\alpha))=q_{0}(\alpha)$ then implies that
$q_{\infty}(\alpha)$ takes value $1$ with probability
$q_{0}(\alpha)=|\braket{\alpha}{\varphi}|^2$ and $0$ with probability
$1-q_{0}(\alpha)$ as expected in a perfect measurement of a non-degenerate
system observable with the $|\alpha\rangle$'s as eigenstates.

The proofs of statements $iii)$ and $iv)$ use the same tools. We start by determining
the rate of convergence to the limiting system state. This turns out to depend
on this limiting state and this is also the clue to statement $iii)$. 
What we have proved so far implies that at some time,
say $n_0$, one of the components, say $q_{n_0}(\gamma)$, will be large, i.e.
close to $1$, so that all other components will be small. 
We can then replace $ (\ref{eq:rs})$ by an approximate linear recursion
relation, namely, for $\alpha \neq \gamma$, 
\begin{equation} \label{eq:rsa}
q_{n+1}(\alpha)=q_{n}(\alpha)\frac{p(i|\alpha)}{ p(i|\gamma)}  
\end{equation}
with probability $p(i|\gamma)$ (if non zero). The proof given above shows again
that this random recursion relation defines a martingale. There is a subtle
point however : this martingale is not bounded anymore and the martingale
convergence theorem does not apply. However, we can rely on a simpler
tool. Defining $l_n \equiv \log q_{n}$ we get, for $\alpha \neq \gamma$,
\begin{equation} \label{eq:lrsa}
l_{n+1}(\alpha)=l_{n}(\alpha)+ \log \frac{p(i|\alpha)}{ p(i|\gamma)}.  
\end{equation}
with probability $p(i|\gamma)$ (if non zero). So $l_{n}(\alpha)-l_{0}(\alpha)$ is
the sum of $n$ independent identically distributed random variables with mean
$-S(\gamma|\alpha)\equiv \sum_i p(i|\gamma) \log p(i|\alpha)/ p(i|\gamma)$.
Remember that for each $\beta$, the collection $p(i|\beta)$, $i\in I$, defines a
probability on $I$, and $S(\gamma|\alpha)$ is nothing but the relative entropy
of $p(i|\gamma)$ with respect to $p(i|\alpha)$, a quantity which is always
non-negative, and in fact strictly positive under the non-degeneracy assumption.
The law of large numbers yields $l_{n}(\alpha)\sim -n S(\gamma|\alpha)
\rightarrow -\infty$, so that $q_{n}(\alpha)$ converges exponentially to $0$
with rate $S(\gamma|\alpha)$. Hence, as soon as one of the components, say
$q(\gamma)$, has become reasonably close to one, with high probability the state
of the system will converge to $\ket{\gamma}$. In this situation, each
measurement on the probe leads to a gain of information on the system state which
in average is given for each component $\alpha \neq
\gamma$ by the relative entropy $S(\gamma|\alpha)$.

By the strong law of large numbers, the previous discussion also implies that if the limit state is $\ket{\gamma}$, the frequency of measurements leading to probe state $\ket{i}$ will converge to $p(i|\gamma)$. By the non-degeneracy hypothesis this fixes the limit pointer state unambiguously. This proves statement $iii)$. In practice, an histogram of all $n_i/n$, the fraction of probes measured in state $\ket{i}$ in a single series of a large number $n$ of repeated measurements, for $i\in I$, will be close to $p(i|\gamma)$ for a single $\ket{\gamma}$, allowing to identify $\ket{\gamma}$.  Then conducting many independent homogeneous series (starting each experiment with the same system state) allows to reconstruct the probabilities $q_0$. Hence the homogeneous repeated measurement scheme is fully equivalent to an ideal Von Neumann measurement. 

To finish the discussion,
note that by the martingale property, knowing the results of probe measurements
up to time $n_0$, the probability to end in pointer state $\ket{\gamma}$ is
exactly $q_{n_0}(\gamma)$, which is close to $1$. The quantity
$1-q_{n_0}(\gamma)$ is the probability to end in another pointer state. It
is also the order of magnitude of the probability that the above discussion
breaks down. This occurs precisely when the random evolution invalidates the linear approximation.
If this happens, it will be likely to happen quickly after $n_0$ because, if
for a long time after $n_0$ the $q_{n}(\alpha)$ remain small, the law of large
numbers implies that they are very likely to decrease exponentially so that
escaping away from the pointer state $\ket{\gamma}$ will get harder and harder. Take
some $\varepsilon > 0$ such that if $1- \varepsilon < q_{n}(\gamma)$ the linear
approximation is good to describe the transition from time $n$
to time $n+1$. Suppose that during a random evolution this condition on
$q_{n}(\gamma)$ remains valid for $n_0 \leq n \leq n_1$. By standard large
deviation theory (Cramer's theorem) if $n_1-n_0$ is large, the probability that,
for a given $\alpha$, $q_{n_1}(\alpha)$ is of order $\varepsilon$ (instead of
being of order $\varepsilon \exp{[-(n_1-n_0)S(\gamma|\alpha)]}$) is estimated
crudely as $\sim \lambda_*^{n_1-n_0}$ for a certain $\lambda^* < 1$ which is the
minimum over $s >0$ of the function $\lambda(s)=\sum_i p(i|\gamma)
\left(\frac{p(i|\alpha)}{ p(i|\gamma)}\right)^s$.

Finally, we emphasize that the (infinite) series of indirect experiments may be viewed as building a measurement apparatus  \cite{mb_prep}. Indeed, the reading of the asymptotic behavior of the frequencies of the probe measurement outcomes allows to register the limit pointer state.\\

{\it Acknowledgements}: We thank Elsa Bernard for sharp informative discussions. This work was in part supported by ANR contract ANR-2010-BLANC-0414. We thank the referee for useful comments and for pointing the interesting refs.\cite{Adler}.


\begin{references}
\bibitem{credit_fig} Fig.\ref{fig:lkb_exp} has been extracted from \cite{lkb_exp} with the authors' and publisher permission (License number: 2695350756236).
\bibitem{limit} Under hypotheses (e.g. weak coupling), one may argue that the discrete setting we are considering admits time-continuous limits coinciding with the models \cite{Gisin-Diosi,GBB}.
\bibitem{point} Point $iii)$ does not seem to have been addressed in \cite{Adler} which only prove the collapse without explaining how to identify the limit pointer state in a given realization. Point $iv)$ shows that the convergence rate depends on the limit pointer state, a fact which contrasts with the usual estimates \cite{Adler}.
\bibitem{martin} Except that real martingales have the property not to cross a given interval an infinite number of times.
\bibitem{period} Due to the periodicity of the evolution operator in \cite{lkb_exp}, the problem of degeneracy is present and circumvented by the fact that the initial state has only negligible components on high energy states.
\end{references}
\end{document}